\documentclass[superscriptaddress,prl,amsmath,amssymb,bibnotes,altaffilletter,twocolumn]{revtex4-1}

\usepackage[latin1]{inputenc}
\usepackage{longtable}  
\usepackage{multirow}
\usepackage{graphicx}
\usepackage{amsmath,amssymb,bbold,bm}
\usepackage{epstopdf}
\usepackage{xcolor}
\usepackage{lipsum}
\usepackage{hyperref}
\usepackage[toc,page]{appendix}
\usepackage{rotating}
\usepackage{blkarray}

\hypersetup{
    colorlinks=true,
    linkcolor=blue,
    filecolor=magenta,      
    urlcolor=cyan,
}
\definecolor{blue}{RGB}{50,0,255}
\definecolor{orange}{RGB}{255,128,0}
\colorlet{blueh}{blue!30}
\usepackage{fancyhdr}

\usepackage{soul}

\begin{document}

\def\nuc#1#2{${}^{#1}$#2}
\def\mee{$\langle m_{\beta\beta} \rangle$}
\def\mnu{$m_{\nu}$}
\def\ml{$m_{lightest}$}
\def\gnu{$\langle g_{\nu,\chi}\rangle$}
\def\mmod{$\| \langle m_{\beta\beta} \rangle \|$}
\def\mb{$\langle m_{\beta} \rangle$}
\def\BBz{$\beta\beta(0\nu)$}
\def\BBm{$\beta\beta(0\nu,\chi)$}
\def\BBt{$\beta\beta(2\nu)$}
\def\nonubb{$\beta\beta(0\nu)$}
\def\twonubb{$\beta\beta(2\nu)$}
\def\BB{$\beta\beta$}
\def\Mz{$M_{0\nu}$}
\def\Mt{$M_{2\nu}$}
\def\MzG{$M^{GT}_{0\nu}$}           
\def\MzF{$M^{F}_{0\nu}$}                
\def\MtG{$M^{GT}_{2\nu}$}           
\def\MtF{$M^{F}_{2\nu}$}                
\def\Gz{$G_{0\nu}$}					
\def\Tz{$T^{0\nu}_{1/2}$}
\def\Tt{$T^{2\nu}_{1/2}$}
\def\Tc{$T^{0\nu\,\chi}_{1/2}$}
\def\Rz{$\Gamma_{0\nu}$}            
\def\Rt{$\Gamma_{2\nu}$}            
\def\ms{$\delta m_{\rm sol}^{2}$}
\def\ma{$\delta m_{\rm atm}^{2}$}
\def\mot{$\delta m_{12}^{2}$}
\def\mtt{$\delta m_{23}^{2}$}
\def\ts{$\theta_{\rm sol}$}
\def\ta{$\theta_{\rm atm}$}
\def\ttwo{$\theta_{12}$}
\def\tot{$\theta_{13}$}
\def\gpp{$g_{pp}$}                  
\def\gA{$g_{A}$}                  
\def\qval{$Q_{\beta\beta}$}                 
\def\be{\begin{equation}}
\def\ee{\end{equation}}
\def\cpKkgy{cnts/(keV kg y)}
\def\cpKkgd{cnts/(keV kg d)}
\def\cpRty{cnts/(ROI t y)}
\def\onecpRty{1~cnt/(ROI t y)}
\def\threecpRty{3~cnts/(ROI t y)}
\def\ppc{P-PC}                          
\def\nsc{N-SC}                          
\def\cosixty{$^{60}Co$}
\def\thttt{$^{232}\mathrm{Th}$}
\def\utte{$^{238}\mathrm{U}$}
\def\mubqkg{$\mu\mathrm{Bq/kg}$}
\def\cusulfate{$\mathrm{CuSO}_4$}
\def\MJ{{\sc Majorana}}             
\def\DEM{{\sc Demonstrator}}             
\def\MJDEMbf{\bfseries{\scshape{Majorana Demonstrator}}}
\def\MJbf{\bfseries{\scshape{Majorana}}}
\def\MJDEMit{\itshape{\scshape{Majorana Demonstrator}}}
\newcommand{\Gerda}{GERDA}
\newcommand{\GF}{\textsc{Geant4}}
\newcommand{\MaGe}{\textsc{MaGe}}

\newcommand{\lanl}{Los Alamos National Laboratory, Los Alamos, NM, USA}
\newcommand{\INR}{Institute for Nuclear Research of the Russian Academy of Sciences, Moscow 117312, Russia}
\newcommand{\SNO}{SNOLAB, Sudbury, ON P3Y 1N2, Canada}
\newcommand{\Osaka}{Research Center for Nuclear Physics, Osaka University, Osaka, Japan}
\newcommand{\IfK}{Institut f\"{u}r Kernphysik, Westfälische Wilhelms-Universität Munster, D-48149 Munster, Germany}
\newcommand{\LBNL}{Nuclear Science Division, Lawrence Berkeley National Laboratory, Berkeley, CA 94720, USA }
\newcommand{\UCB}{Department of Physics, University of California, Berkeley, CA 94720, USA }
\newcommand{\JINR}{Joint Institute for Nuclear Research (JINR) Joliot-Curie 6, 141980, Dubna, Moscow Region, Russia}
\newcommand{\NIST}{National Institute of Standards and Technology, 100 Bureau Dr, Gaithersburg, MD 20899, USA}
\newcommand{\JSC}{JSC `State Scientific Center Research Institute of Atomic Reactors', Dimitrovgrad, 433510, Russia }
\newcommand{\UW}{Center for Experimental Nuclear Physics and Astrophysics, and Department of Physics, University of Washington, Seattle, WA 98195, USA}
\newcommand{\Carlton}{Carleton University 1125 Colonel By Drive Ottawa, K1S 5B6, Canada}
\newcommand{\UNC}{Department of Physics and Astronomy, University of North Carolina, Chapel Hill, NC 27599, USA}
\newcommand{\TUNL}{Triangle Universities Nuclear Laboratory, Durham, NC 27708, USA}

\title{Results from the Baksan Experiment on Sterile Transitions (BEST) }
\affiliation{\INR}
\affiliation{\SNO}
\affiliation{\Osaka}
\affiliation{\lanl}
\affiliation{\IfK}
\affiliation{\LBNL}
\affiliation{\UCB}
\affiliation{\JINR}
\affiliation{\NIST}
\affiliation{\JSC}
\affiliation{\UW}
\affiliation{\Carlton}
\affiliation{\UNC}
\affiliation{\TUNL}

\author{V.V.~Barinov}\affiliation{\INR}
\author{B.T.~Cleveland}\affiliation{\SNO}
\author{S.N.~Danshin}\affiliation{\INR}
\author{H.~Ejiri}\affiliation{\Osaka}
\author{S.R.~Elliott}\affiliation{\lanl}
\author{D.~Frekers}\affiliation{\IfK}
\author{V.N.~Gavrin}\email[]{gavrin@inr.ru}\affiliation{\INR}
\author{V.V.~Gorbachev}\affiliation{\INR}
\author{D.S.~Gorbunov}\affiliation{\INR}
\author{W.C.~Haxton}\affiliation{\LBNL}\affiliation{\UCB}
\author{T.V.~Ibragimova}\affiliation{\INR}
\author{I.~Kim}\affiliation{\lanl}
\author{Yu.P.~Kozlova}\affiliation{\INR}
\author{L.V.~Kravchuk}\affiliation{\INR}
\author{V.V.~Kuzminov}\affiliation{\INR}
\author{B.K.~Lubsandorzhiev}\affiliation{\INR}
\author{Yu.M.~Malyshkin}\affiliation{\INR}
\author{R.~Massarczyk}\affiliation{\lanl}
\author{V.A.~Matveev}\affiliation{\JINR}
\author{I.N.~Mirmov}\affiliation{\INR}
\author{J.S.~Nico}\affiliation{\NIST}
\author{A.L.~Petelin}\affiliation{\JSC}
\author{R.G.H.~Robertson}\affiliation{\UW}
\author{D.~Sinclair}\affiliation{\Carlton}
\author{A.A.~Shikhin}\affiliation{\INR}
\author{V.A.~Tarasov }\affiliation{\JSC}
\author{G.V.~Trubnikov}\affiliation{\JINR}
\author{E.P.~Veretenkin}\affiliation{\INR}
\author{J.F.~Wilkerson}\affiliation{\UNC}\affiliation{\TUNL}
\author{A.I.~Zvir}\affiliation{\JSC}

\begin{abstract}

The Baksan Experiment on Sterile Transitions (BEST) was designed to investigate the deficit of electron neutrinos, $\nu_{e}$, observed in previous gallium-based radiochemical measurements with
high-intensity neutrino sources, commonly referred to as the \textit{gallium anomaly}, which could be interpreted as evidence for oscillations between $\nu_e$ and sterile neutrino ($\nu_s$) states. A 3.414-MCi \nuc{51}{Cr} $\nu_e$ source was placed at the center of two nested Ga volumes and measurements were made of the production of \nuc{71}{Ge} through the charged current reaction, \nuc{71}{Ga}($\nu_e$,e$^-$)\nuc{71}{Ge}, at two average distances. The measured production rates for the inner and the outer targets respectively are  ($54.9^{+2.5}_{-2.4}(\mbox{stat})\pm1.4 (\mbox{syst})$) and  ($55.6^{+2.7}_{-2.6}(\mbox{stat})\pm1.4 (\mbox{syst})$) atoms of \nuc{71}{Ge}/d. The ratio ($R$) of the measured rate of \nuc{71}{Ge} production at each distance to the expected rate from the known cross section and experimental efficiencies are  $R_{in}=0.79\pm0.05$  and $R_{out}= 0.77\pm0.05$. The ratio of the outer to the inner result is 0.97$\pm$0.07, which is consistent with unity within uncertainty. The rates at each distance were found to be similar, but 20-24\% lower than expected, thus reaffirming the anomaly. These results are consistent with $\nu_e \rightarrow \nu_s$ oscillations with a relatively large $\Delta m^2$ ($>$0.5~eV$^2$) and mixing sin$^2 2\theta$ ($\approx$0.4).
\end{abstract}


\date{\today}
\maketitle




The possibility of the existence of light sterile neutrinos ($\nu_s$) is presently a major field of inquiry. The literature on this topic is extensive but has been summarized well in a number of recent reviews~\cite{abazajian2012light,Gariazzo2015,Giunti2019,Boser2020,Diaz2020,seo2020review,dasgupta2021sterile}. Much of the evidence for $\nu_s$'s comes from oscillation experiments that search for the conversion of an active neutrino into a sterile state. 

\begin{figure}[thb]
    \centering
    \includegraphics[width=0.8\columnwidth]{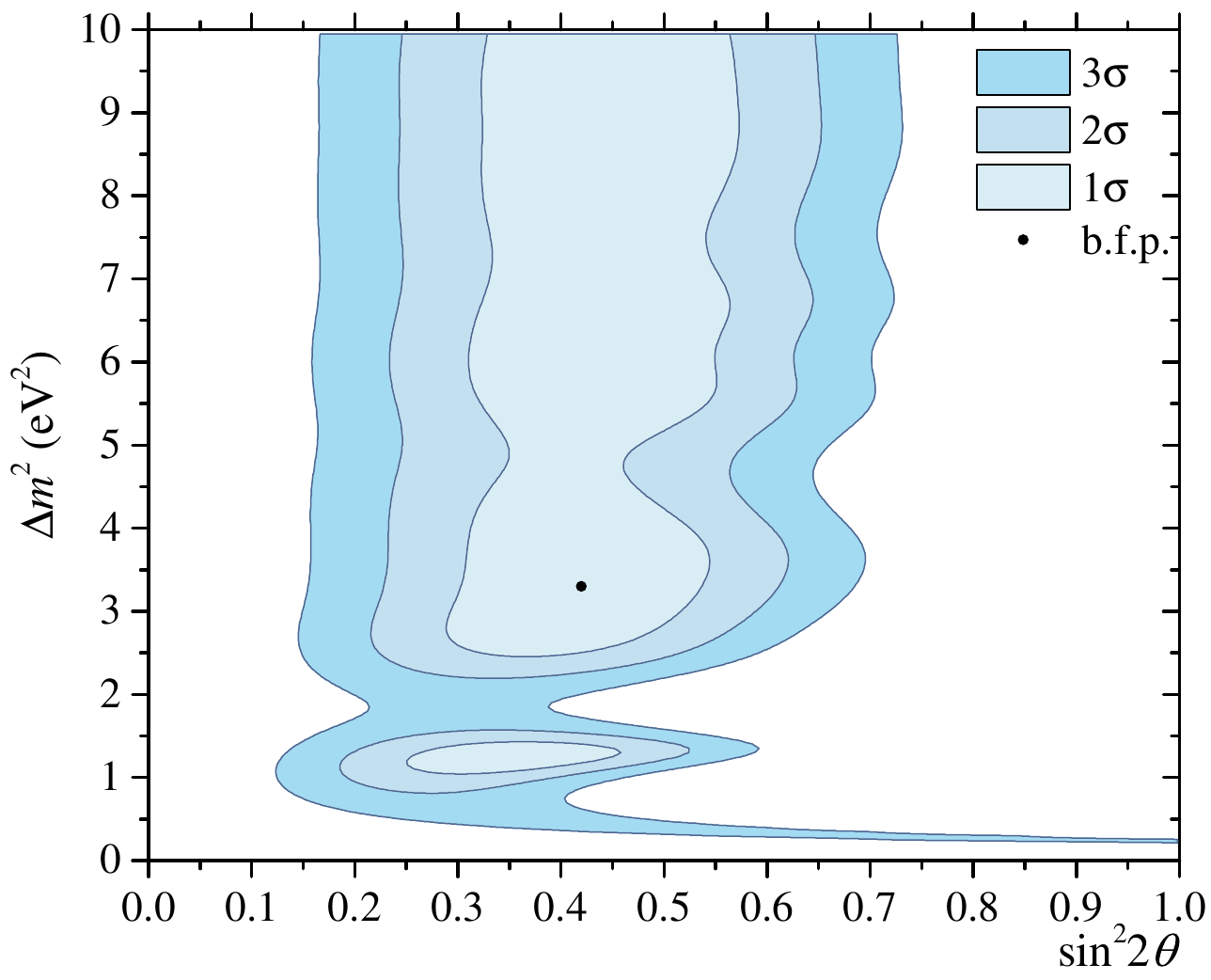}
 \includegraphics[width=0.9\columnwidth]{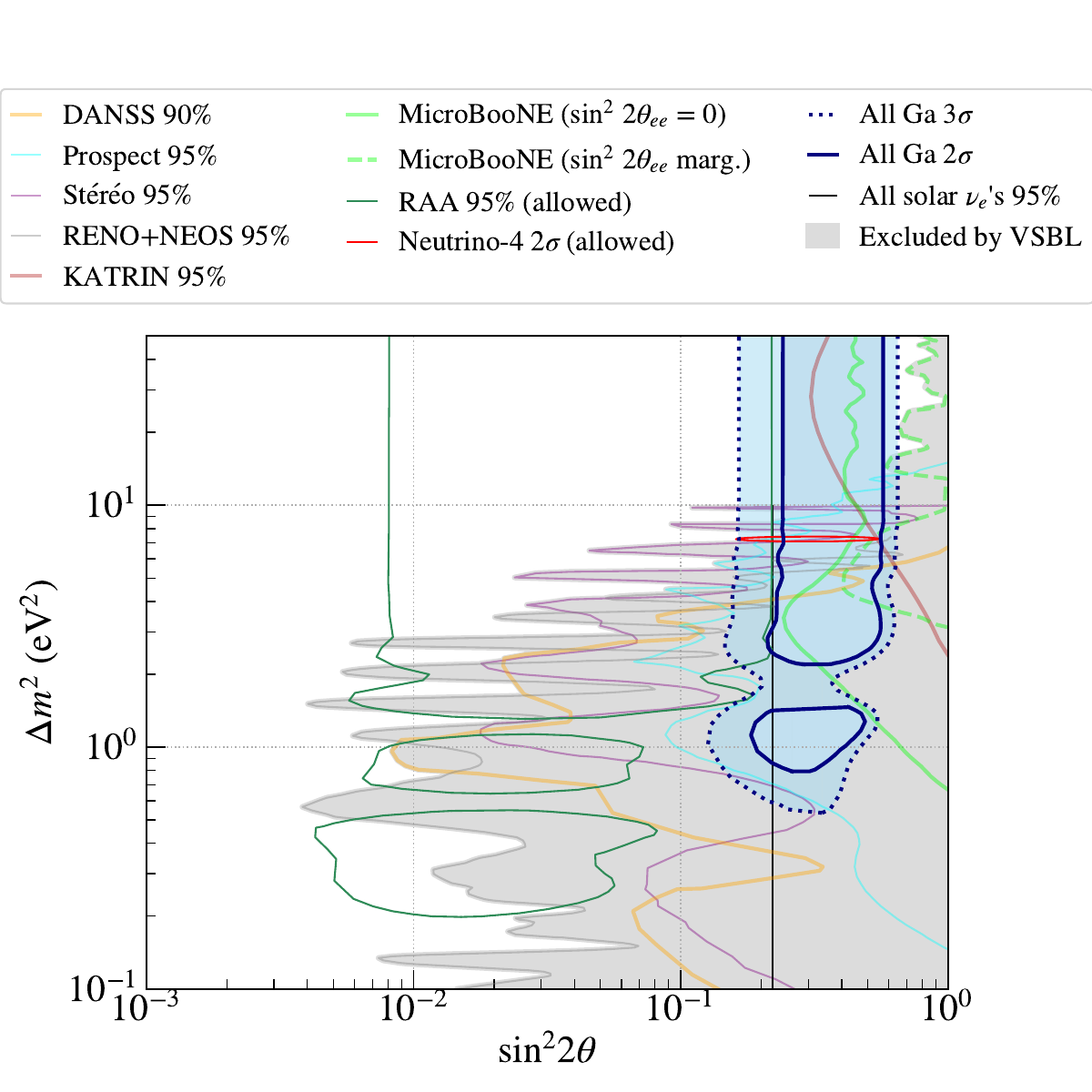}
    \caption{Top: Exclusion for the results from BEST. The best fit point (b.f.p.) is for $\Delta m^2$ = 3.3 eV$^2$ and sin$^2 2\theta$ = 0.42. Bottom: Exclusion contours of all Ga anomaly experiments: two GALLEX, two SAGE and two BEST results. The blue solid line and the blue dotted line shows the 2$\sigma$ and 3$\sigma$ confidence level respectively. The figure also presents the exclusion contours  from Prospect~\cite{prospect2021}, DANSS~\cite{danss2020}, St\'{e}r\'{e}o~\cite{stereo2020}, KATRIN~\cite{katrin2020}, the combined analysis of RENO and NEOS data~\cite{atif2020search}, reactor anti-neutrino anomalies~(RAA)~\cite{chooz2011reactor} allowed region, interpretations of the MicroBooNE result for the oscillation hypothesis with fixed mixing angle~(sin$^22\theta$) and profiled over the angle~\cite{arguelles2021microboone}, and the model-independent 95\% upper bound on sin$^2 2\theta$ from all solar neutrino experiments~\cite{neutrino4_2021}. The 2$\sigma$ allowed region of Neutrino-4~\cite{serebrov2020analysis} is also presented and the grey shading represents the merged exclusion of the very short baseline (VSBL) null results.}
    \label{fig:exclusion}
\end{figure}

The SAGE~\cite{Abdurashitov2009PRC} and GALLEX~\cite{Kaether2010} radiochemical  experiments detected neutrinos from the Sun through the charged-current reaction \nuc{71}{Ga}($\nu_e$,e$^-$)\nuc{71}{Ge}. The SAGE method (GALLEX) exposed a large mass of Ga metal, 30-50 t, (GaCl$_3$-HCl solution, 30.3 t Ga) to the Sun for about a month and then chemically extracted the radioactive \nuc{71}{Ge} atoms ($\tau_{1/2}$=(11.43$\pm$0.03)~d~\cite{Hampel1985}), mixed the Ge with a proportional counter gas, and counted the decaying \nuc{71}{Ge} in a low-background system. Both collaborations followed up the solar neutrino studies with strong radioactive electron-capture sources to confirm their sensitivity to interactions with $\nu_e$ from the Sun. These experiments, using \nuc{51}{Cr}~\cite{Abdurashitov1999,gallex1998cr51} and \nuc{37}{Ar}~\cite{Abdurashitov2006} placed at the center of their Ga targets, found a \nuc{71}{Ge} production rate of 0.87$\pm$0.05 of that expected~\cite{Abdurashitov2009PRC}. This led to extensive studies of the cross section~\cite{Barinov2018,Frekers2011,Frekers2013,Frekers2015,Semenov2020}, the extraction efficiency, and counting efficiencies~\cite{gallex1998cr51,Abdurashitov2002} by both collaborations and a number of outside interested groups~\cite{Kaether2010,Giunti2012,Kostensalo2019}. This discrepancy between the expected and measured rates defines the \textit{gallium anomaly} and has been interpreted in the context of $\nu_e \rightarrow \nu_s$ oscillations~\cite{LAVEDER2007344}. Although the statistical evidence for a deviation from expectation is modest, about 2-3$\sigma$,  it has persisted motivating the need for further investigation. Furthermore, given the simplicity of the electron-capture neutrino energy spectrum and the well-known cross section ($\sigma$) at these low energies, this is an effective technique to search for $\nu_s$'s. There have been numerous searches for $\nu_s$ sensitive to the Ga anomaly parameter range. We present a summary in the bottom panel of Fig.~\ref{fig:exclusion}.

The use of an electron capture $\nu_e$ source is a powerful technique to search for $\nu$ oscillations. The $\nu_e$ spectrum from \nuc{51}{Cr} is simple, being comprised of a dominant component near 750~keV and a sub-dominant component near 430~keV. It is a well-understood spectrum relying on well-known nuclear and atomic physics parameters. The \nuc{51}{Cr} isotope (27.704$\pm$0.004~d) emits $\nu_e$'s at four energies; 747~keV (81.63\%), 427~keV (8.95\%), 752~keV (8.49\%) and 432~keV (0.93\%). 

\begin{figure*}[ht]
 \centering
 \includegraphics[width=15cm]{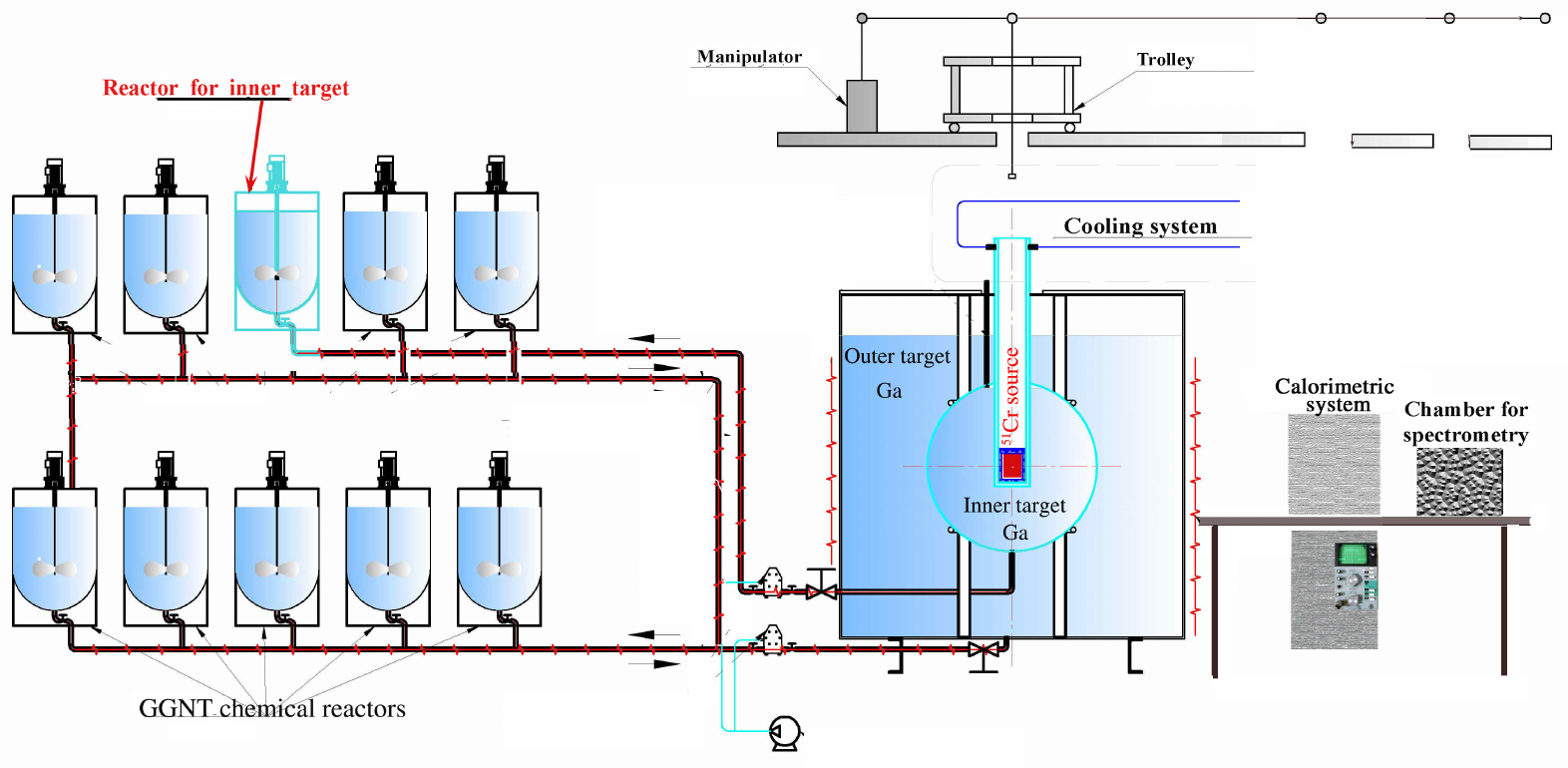}
 \caption{The Ga target and extraction piping diagram also indicating the source handling apparatus.}
 \label{fig:TargetGeometry}
\end{figure*}

The previous source measurements used a single target and, therefore, required comparison of a measured rate to a theoretical expectation. The Baksan Experiment on Sterile Transitions (BEST) was designed as a two-distance oscillation experiment. The experimental concept is depicted in Fig.~\ref{fig:TargetGeometry}. An inner spherical volume, with diameter 133.5~cm, contains (7.4691$\pm$0.0631)~t of Ga. An outer cylindrical volume (234.5~cm high, 218~cm diam.) contains (39.9593$\pm$0.0024)~t of Ga. The $^{51}$Cr source was placed at the center irradiating both volumes simultaneously, permitting the production rate  of \nuc{71}{Ge} to be measured at two different distances. After exposure, the Ga was pumped to reactors for the extraction chemistry. Detailed discussion of the experimental operations, efficiencies and uncertainties can be found in Ref.~\cite{BESTPRC}.

The active core of the source consisted of 26 irradiated Cr disks, placed into a stainless-steel cylinder with a radius of 4.3~cm and height 10.8~cm, 
was shielded for radiation safety within a tungsten alloy of thickness of $\approx$30~mm. The source was manufactured by irradiating 4~kg of \nuc{50}{Cr}-enriched metal for 100~d in a reactor at the State Scientific Center Research Institute of Atomic Reactors, Dimitrovgrad, Russia. The source was delivered to the Baksan Neutrino Observatory (BNO) on July 5, 2019 and was placed into the two-zone target at 14:02 that same day and this is our chosen reference time for the source strength.  The activity ($A$) at the reference time is (3.414$\pm$0.008) MCi. A full description of the source and the calorimetric measurements of its intensity can be found in Refs.~\cite{Kozlova2020,Gavrin_2021}. 

Twenty extractions, 10 from each volume, were conducted between July 15 and Oct. 13, 2019.  The Ga metal was kept molten by maintaining the temperature between 30.0~C and 30.5~C above the 29.8~C melting point. At the start of each exposure, $\approx$175~$\mu$g of Ge carrier was added to each Ga volume. Each exposure lasted approximately 9~d and counting of the sample commenced approximately 24~h after each extraction. At the end of the exposure, the carrier Ge and any produced \nuc{71}{Ge} was extracted using the procedure described in Ref.~\cite{CLEVELAND201541}. The process ensured the independent extraction of \nuc{71}{Ge} atoms from each zone of the Ga target. The gas germane was synthesized, mixed with Xe, and inserted into small ($\sim$0.6~cm$^3$), low-background  proportional counters. The counters were installed into a NaI well volume of one of the two counting systems. The counting duration varied from 60 to 150~days. The irradiations were scheduled to maximize the number of extracted  \nuc{71}{Ge} atoms. The first extraction's counting times were shorter due to the limited number of working counters. The shorter counting time had little effect on the number of measured \nuc{71}{Ge} decays, but the statistical uncertainty was increased due to the lower statistical determination of the counter background.


Two 8-channel data acquisition systems were used~\cite{Shikhin2017,Shikhin2018,Abdurashitov1999a}. Pulses from the proportional counters were digitized at 1~GHz with a bandwidth of 100~MHz and a rise time of 3.5~ns. The energy range of the 8-bit digitizer was 0.37-15~keV. The energy range of the NaI counters was 60-3000~keV. Data were collected for each individual event, recording the time of occurrence. 

The digitized pulse shapes were analyzed for energy and rise time~\cite{Abdurashitov1999a}. The measure of energy is the integral of the pulse waveform for 800~ns after pulse onset. The peak position for each counter is based on routine periodic calibrations with \nuc{55}{Fe}. Auger electrons and x rays from \nuc{71}{Ge} decay will produce point-like ionization within the gas resulting in a short rise time compared to an extended ionization trail arising from Compton electrons or $\beta$ particles. Thus the pulse rise time ($T_N$) can be used to eliminate background and was determined by a functional fit to the waveform~\cite{Elliott1990}.  After counting of the samples from the Cr experiment was completed in fall of 2020, measurements of the counting efficiency were made for each counter used in the experiment. Two different techniques and two different isotopes were employed: \nuc{37}{Ar} to measure volume efficiency, and \nuc{71}{Ge} to measure the L- and K-peak efficiencies and the $T_N$ acceptance for each counter. An upper limit for $T_N$ consistent with point-like events was determined such that 96\% of the \nuc{71}{Ge} events were accepted~\cite{Abdurashitov2006}. The volume efficiencies of all counters used in the experiment were directly measured with \nuc{37}{Ar}. The calculated counting efficiency using the measured pressure, GeH$_4$ fraction, and \nuc{37}{Ar} volume efficiency was determined for each extraction. The total uncertainty in these calculated efficiencies is $\pm$1.1\%. The total efficiency varies for each extraction and is the product of the live time factor, counting efficiency with analysis cuts, extraction and synthesis efficiency, and a factor due to the \nuc{51}{Cr} half-life during the exposure is typically (10.0$\pm$0.3\%)~\cite{BESTPRC}, where  systematic uncertainties are included.

\begin{figure}[thb]
 \centering
 \includegraphics[width=.8\columnwidth]{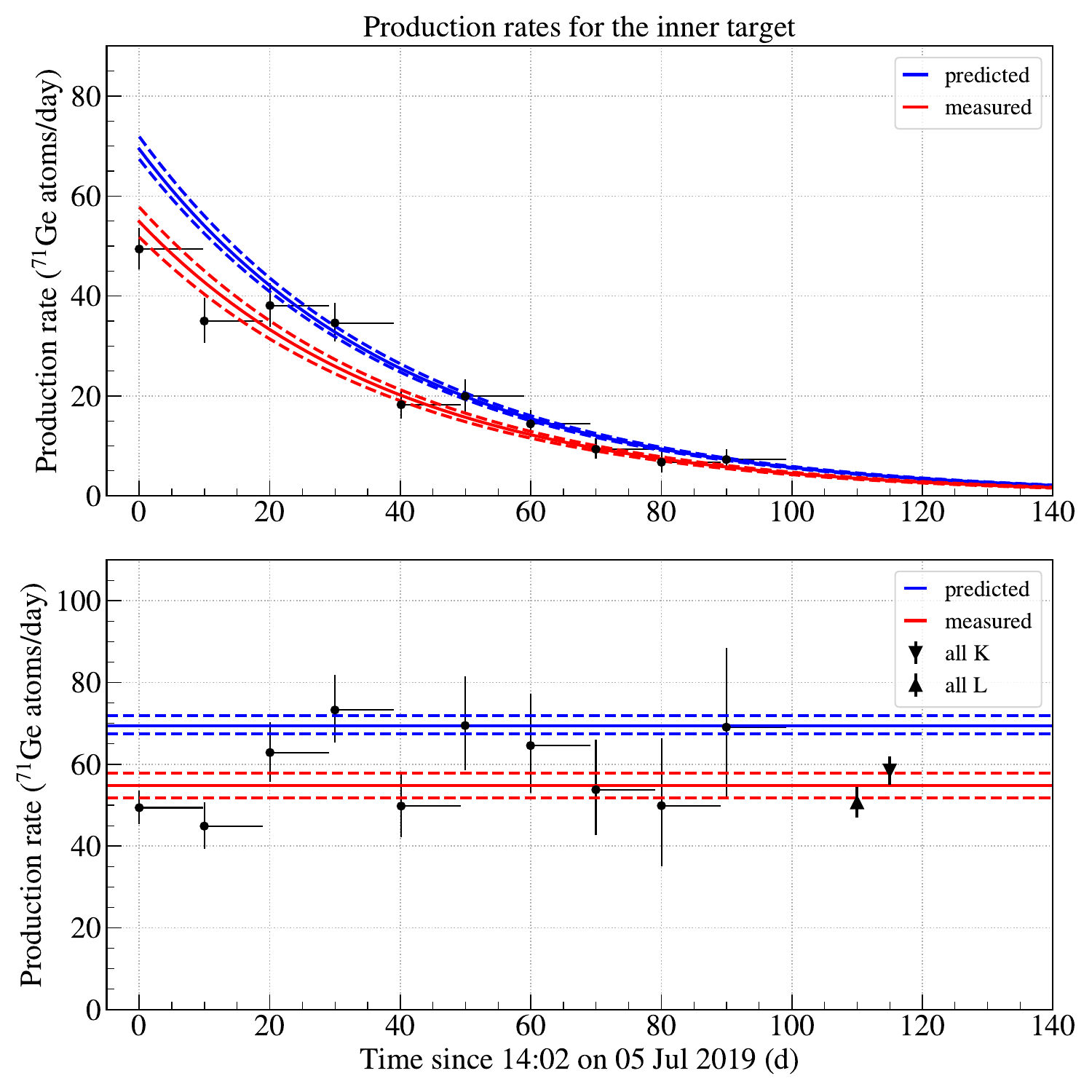}
 \includegraphics[width=.8\columnwidth]{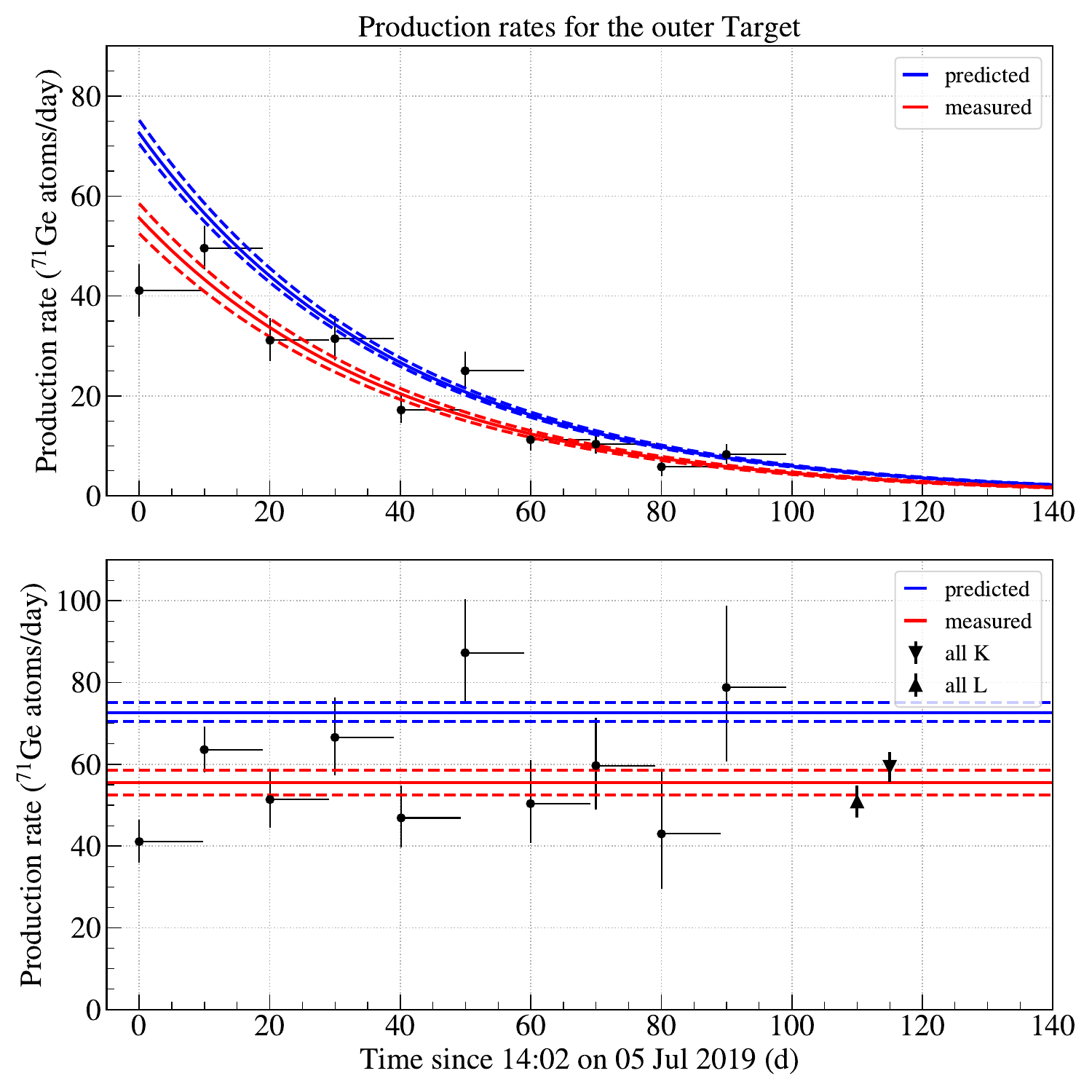}
 \caption{Top: the measured K+L peak rates of the inner target volume; Middle-Top: normalizes the production rate to the reference time, the combined results for events in the the L and K peaks are shown. The blue (red) region represents the predicted (measured) production rate. Middle-Bottom: Similar to the Top panel but for the outer volume. Bottom: Similar to the Middle-Top panel but for the outer volume. The dotted lines enclose the $\pm1\sigma$ uncertainty regions. For all panels, the horizontal lines indicate the exposure duration with the likelihood fit results plotted at the start of exposure.}
 \label{fig:KLPeakFits}
\end{figure}

The likelihood fits to the time distribution of the candidate events were performed as in Ref.~\cite{Abdurashitov1999}. This analysis includes a \nuc{71}{Ge} contribution with its 11-d half-life and a constant background rate. For joint fits of all extractions, the decay of the \nuc{51}{Cr} source was taken into account. Table~\ref{tab:CountResults} presents a summary of the K+L fit results for each extraction for the inner and outer volumes. Additionally, a  combined fit for each is given. Figure~\ref{fig:KLPeakFits} shows the K+L production rate fits for the two volumes indicating the resulting production rate at the reference time. Two independent analyses were pursued and both obtained similar results to within about 2\%. This difference is due to minor event-selection differences at the edges of the selection borders in energy and rise time. This difference is accounted for by the estimated systematic uncertainties in the efficiencies for those cuts. All efficiencies are accounted for each extraction individually.

For the likelihood fits, if the $^{71}$Ge half-life is allowed to float, the result is 11.05$\pm$0.72~d (11.11$\pm$0.69~d)  for the inner (outer) target data agreeing well with the known half-life. If the $^{51}$Cr half-life is allowed to float the result is 31.55$\pm$2.89~d (30.97$\pm$3.90~d)  for the inner (outer) target data agreeing well with the known half-life.

\begin{table*}[t]
\caption{A summary of the likelihood fits for the production rate from each extraction, the combined fit of all extractions, and the predicted production rate. The 2$^{nd}$ and 6$^{th}$ columns are the total number of energy and rise-time selected candidates for $^{71}$Ge decay. The 3$^{rd}$ and 7$^{th}$ columns are the number of candidates that fit to $^{71}$Ge. The fit background values can be calculated by subtracting columns 3 from 2 or 7 from 6, respectively. The 4$^{th}$ and 8$^{th}$ columns are the number of events assigned to production by $^{51}$Cr after contributions from carryover and solar neutrino production are subtracted. Columns 5 and 9 are the resulting production rates quoted at the reference time. The quoted measurement uncertainties are statistical.}
\label{tab:CountResults}
\begin{tabular}{c||cccc||cccc}
\hline  \hline
				& \multicolumn{4}{c||}{Inner Volume}								&  \multicolumn{4}{c}{Outer Volume} \\
\hline
Exposure			& K+L		& Number fit		& \nuc{51}{Cr}		& Production			& K+L		& Number fit		& \nuc{51}{Cr}		& Production		\\
Dates (DoY)		&Candidates	& to \nuc{71}{Ge}	& Production		&Rate (Atoms/d)		&Candidates	& to \nuc{71}{Ge}	& Production		& Rate (Atoms/d) 	\\
\hline
186.585-196.376	&   180		&   176.3			& 175.5			& 49.4$^{+4.2}_{-4.0}$	& 181		&  133.4			& 129.6			&$41.1^{+5.3}_{-5.2}$\\
197.362-206.372	& 129		&	111.5		&107.7			& 44.9$^{+5.9}_{-5.6}$	& 174		&	163.8		&	158.6		&63.6$^{+5.7}_{-5.5}$\\
207.282-216.374	&	132		&	117.6		& 115.4			& 62.9$^{+7.4}_{-7.1}$	&	116		&	92.5			&	88.2			&51.4$^{+7.3}_{-6.9}$\\
217.286-226.371	&	93		&	87.3			& 85.6			& 73.3$^{+8.6}_{-8.0}$	&	98		&	82.3			&	78.9			&66.6$^{+9.8}_{-9.2}$\\
227.258-236.458	&	134		&	60.2			& 58.4			& 49.8$^{+8.2}_{-7.7}$	&	120		&	64.0			&	59.5			&46.9$^{+7.9}_{-7.2}$\\
237.342-246.369	&	81		&	48.8			& 47.7			& 69.5$^{+12.0}_{-11.0}$	&	97		&	62.3			&	59.3			&87.3$^{+13.2}_{-12.3}$\\
247.243-256.368	&	91		&	45.0			& 43.9			&64.6$^{+12.6}_{-11.6}$	&	69		&	38.0			&	34.4			&50.4$^{+10.6}_{-9.6}$\\
257.241-266.369	&	59		&	33.6			& 32.4			&53.8$^{+12.2}_{-11.0}$	&	68		&	43.4			&	39.2			&59.7$^{+11.7}_{-10.8}$\\
267.240-276.369	&	106		&	23.7			& 22.7			&49.9$^{+16.5}_{-14.9}$	&	66		&	20.2			&	17.0			&43.0$^{+15.3}_{-13.5}$\\
277.201-286.367	& 88			&	25.2			&24.3			&69.1$^{+19.4}_{-17.3}$	&	81		&	31.8			&	28.0			&78.8$^{+20.0}_{-18.1}$\\
\hline 
Combined			& 1093		&	724.0		&	708.2		&54.9$^{+2.5}_{-2.4}$	&	1069		&	738.8		&	699.8		&55.6$^{+2.7}_{-2.6}$\\
Predicted			&			&				&				&69.41$^{+2.5}_{-2.0}$ 	&			&				&				&72.59$^{+2.6}_{-2.1}$\\
\hline\hline
\end{tabular}
\end{table*}

During each extraction a small fraction of the production is due to solar neutrinos. The measured solar neutrino capture rate is (66.1$\pm$3.1)~SNU~\cite{Abdurashitov2009PRC}\footnote{1 SNU corresponds to one neutrino capture per second in a target that contains 10$^{36}$ atoms of the neutrino absorbing isotope, in our case \nuc{71}{Ga}} and typically results in about 0.51 (3) counts per extraction attributed to the K+L counts for the inner (outer) target. Due to the inefficiency of the extraction, there are also some \nuc{71}{Ge} atoms that \textit{carryover} from one extraction to the next. Typically this is about 1 count for each volume. Both of these effects were taken into account, extraction by extraction.

The systematic uncertainties have been estimated from auxiliary tests. The chemical extraction efficiency is typically about 95\% with an uncertainty of $\pm$1.6\%. The  summed K+L peak counting efficiency is typically about 70\% with an uncertainty of -1.8/+2.0\%. There are small uncertainties due to the Rn cut (-0.05\%), the solar neutrino correction ($\pm$0.20\%), and the carryover correction ($\pm$0.04\%). The total systematic uncertainty is estimated to be -2.5/+2.6\%. Note that the uncertainty in the extraction efficiency has been greatly reduced as compared to Ref.~\cite{Abdurashitov2006}. This is due to the use of mass spectrometry to determine with high accuracy the efficiency of extraction of minute quantities of Ge from a large mass of Ga~\cite{CLEVELAND201541}. The details of the systematic uncertainties are described in Ref~\cite{BESTLong}.

The cross section has to be calculated from nuclear physics input and when the original \textit{Ga anomaly} was observed, there was concern that the transition strengths to excited states were not fully understood. Bahcall~\cite{Bahcall1997} derived the ground state contribution from the \nuc{71}{Ge} half-life, but the excited state contributions were estimated from charge exchange (i.e. (p,n)) reactions. For the central value, Bahcall used the best estimate of the transition strength values to the excited states with an estimated uncertainty to be the change in $\sigma$ (-1.6/+2.8\%), if one ignores the excited states. The charge exchange data has been improved by recent work~\cite{Frekers2011,Frekers2013,Frekers2015}  indicating that they are not the cause of the discrepancy. However, the excited-state contribution uncertainty is critical because the (p,n) measurements have a significant cancellation between the Gamow-Teller and tensor matrix elements resulting in an underestimate of the transition strengths~\cite{HAXTON1998110}.  Kostensalo \textit{et al.}~\cite{Kostensalo2019} used a nuclear shell model calculation to avoid the (p,n) measurement drawback. The paper of Semenov \textit{et al.}~\cite{Semenov2020} reproduces Bahcall's approach but uses modern values for the transition strengths~\cite{ALANSSARI2016}. The Semenov \textit{et al.} and Kostensalo \textit{et al.} results differ by about 4\%, which  is about 2-3 times larger than the uncertainty estimated for each. Interestingly, the original Bahcall number is half way between these two results with an uncertainty that encompasses both. We therefore use the Bahcall $\sigma$ value and the associated conservative uncertainties from his estimate: $(5.81^{+0.21}_{-0.16})\times10^{-45}$~cm$^2$.


The survival probability at a distance $d$ for two-component oscillation for a given $\nu$ energy ($E_{\nu}$) is 
\begin{equation}
P_{ee}(d) = 1 - \mbox{sin}^22\theta \mbox{sin}^2 \left(1.27 \frac{\Delta m^2[\mbox{eV}^2] d[\mbox{m}]}{E_{\nu}[\mbox{MeV}]} \right).
\end{equation}
where $\Delta m^2$ is the difference of the masses squared between the two neutrino species and $\theta$ is the angle that defines the mixing between them. The capture rate ($r$) can be written
\begin{equation}
r = \int_{V} F \sum_{i=1}^{4}(f_iP_{ee}^i) \sigma n d\vec{x},
\end{equation}
\noindent where $F$ is the flux of $\nu_e$, $P_{ee}^i$ is the oscillation survival probability for the $i^{th}$ neutrino branch with branching fraction $f_i$, $\sigma$ is the cross section, $n$ is the \nuc{71}{Ga} number density 
($(2.1001\pm0.0008)\times10^{22}$/cm$^3$) 
and the integral is calculated over the target volume ($V$). With $A$ as the source activity and $d$ the distance between emission and absorption of the $\nu_e$, this can be written
\begin{equation}
\label{eqn:OscillationRate}
r = \frac{n \sigma A}{4\pi}\int_{V} \frac{\sum_{i}(f_iP_{ee}^i(d))}{d^2} d\vec{x}.
\end{equation}
The integral is calculated by Monte Carlo due to the complexity of the target geometry. The average path length $<L>$ of a neutrino through the target is given by the integral when $P_{ee}=1$. The average path lengths for the BEST volumes are $<L>_{in}=(52.03\pm0.18$)~cm and $<L>_{out}=(54.41\pm0.18$)~cm. The uncertainties on these numbers are dominated by dimensional uncertainties of the apparatus. 



For $n=1,\dots,N$ experiments (the two BEST volumes are treated separately), oscillation parameters are estimated by a global minimization of 
\begin{equation}
\chi^2(\Delta m^2,\sin^22\theta)=({\bf r^{meas}}-{\bf r^{calc}})^T{\bf V}^{-1} ({\bf r^{meas}}-{\bf r^{calc}})\label{eqn:chisquare}
\end{equation}
with the ${\bf r^{meas}}$ (${\bf r^{calc}}$) is the vector of the measured (calculated) rates with $r_i^{calc}(\Delta m^2;\mbox{sin}^22\theta)$ and the covariance matrix
%
%
%
\begin{equation}
V_{nk}=\delta_{nk}\varepsilon_n^2 + \varepsilon_{CS}^n \times \varepsilon_{CS}^k
\end{equation}
where $\varepsilon_n^2=\varepsilon_{n,stat}^2+\varepsilon_{n,syst}^2$ are
uncorrelated uncertainties comprised of statistical and systematic measurement
uncertainties, and  $\varepsilon_{CS}^n$ represent the correlated uncertainties of $\sigma$~\cite{Fogli2002}.

The calculation of the confidence level contours corresponding to a $\Delta\chi^2 = \chi^2 - \chi^2_{min}$ with two degrees of freedom: $\Delta\chi^2$ = 2.30, 6.18, 11.83 for 68.27\% (1$\sigma$), 95.45\% (2$\sigma$) and 99.73\% (3$\sigma$) C.L., respectively. The two BEST results for the measured to expected ratios are $R_{out}$ = 0.77$\pm$0.05 and $R_{in}$ = 0.79$\pm$0.05. The results from SAGE are $R_{Cr}=0.95\pm0.12$~\cite{Abdurashitov1999}, $R_{Ar}=0.79^{+0.09}_{-0.10}$~\cite{Abdurashitov2006} 
 and for GALLEX are $R_{Cr1}=0.95\pm0.11$ and $R_{Cr2}=0.81\pm0.11$~\cite{Kaether2010,Altmann2005}.

Figure~\ref{fig:exclusion} shows the allowed $\Delta m^2$ - sin$^2 2\theta$ parameter space assuming that $\nu_e \rightarrow \nu_s$ oscillations is the origin of the \textit{gallium anomaly}. The best fit for BEST only data is $\Delta m^2 = 3.3$~eV$^2$ and sin$^2 2\theta$ = 0.42. Including all the Ga data, the result is $\Delta m^2 = 1.25$~eV$^2$ and sin$^2 2\theta$ = 0.34. As shown in the figure, the allowed ranges for these parameters are large, however, due to the broadness of the minimum.


The $\nu_e-\nu_s$ oscillation parameter space minimum (Fig.~\ref{fig:exclusion}) is very broad and gradual with very small $\chi^2$ difference between the two best fit points. Because the values for $R$ are similar for the two volumes, the deduced oscillation length is similar to, or smaller, than the volumes' dimensions. As a result, the acceptable $\Delta m^2$  range extends above a lower limit. As a consequence, it is not well determined and the results are consistent with values above about 0.5~eV$^2$. The large deviation of the $R$'s from 1 drives the mixing angle to a large value within an extended range. This description is similar to the previous Ga results and hence, given the broad minimum, the difference in parameter values at the minima points is inconsequential. 

Because the measured $R$'s for the two volumes are similar, an alternative explanation for the results could be an overall error in $\sigma$ or efficiency. Since the observed $R$'s would require a smaller $\sigma$ than the ground state contribution alone, some fundamental misunderstanding of the nuclear or atomic physics would be necessary for a reduced $\sigma$ to resolve the Ga anomaly. Given the known  \nuc{71}{Ge} decay rate,  $\sigma$ to the ground state is assumed to be well determined and the inclusion of excited state contributions cannot decrease $\sigma$. 

An error in the efficiency also cannot be ruled out but the experimental procedures have been verified extensively over the past two decades. Many aspects of BEST have been double-checked, including the Ga target masses, the extraction efficiency, the source strength, the source placement, the counting efficiency and the counting system operation. No cause for concern was found.   

After the BEST measurements the Ga anomaly looks more pronounced; the weighted average value of the neutrino capture rate relative to the expected value for all Ga experiments is 0.80$\pm$0.05, accounting for the correlated uncertainty for $\sigma$.  If one ignores the excited state contribution to $\sigma$, the value would increase to 0.84$\pm$0.04, but still be significantly below 1. The hypothesis of $\nu_e \rightarrow \nu_s$ oscillations with a large mass difference ($\Delta m^2 \gtrsim 0.5$~eV$^2$) and large mixing angle (sin$^2 2\theta \approx0.4$) is consistent with these results. A future source experiment with a smaller inner volume might be considered, but the required source strength would be challenging.

We thanks to V. A. Rubakov for the interest, stimulation and fruitful
discussions. This work is supported by former Federal Agency for
Scientific Organizations (FANO till 2019) of Russian Federation, Ministry
of Science and Higher Education of Russian Federation under agreement no.
14.619.21.0009 (unique project identifier no.RFMEFI61917X0009), State
Corporation ROSATOM, and the Office of Nuclear Physics of the US
Department of Energy.

\bibliography{BESTReferences}

\end{document}